\documentclass[%
 jor,
 jor,
 amsmath,amssymb,
preprint,
author-numerical,
longbibliography
]{revtex4-1}

\usepackage{epsfig,verbatim,hyperref,color}

\begin{document}

\preprint{AIP/123-QED}

\title{A Small-gap Effective-Temperature Model of Transient Shear Band Formation During Flow}

\author{Adam R. Hinkle}
\affiliation{Department of Materials Science \& Engineering, Johns Hopkins University, Baltimore, MD 21218 USA}

\author{Michael L. Falk}
\affiliation{Department of Materials Science \& Engineering, Johns Hopkins University, Baltimore, MD 21218 USA}
\affiliation{Department of Physics \& Astronomy}
\affiliation{Department of Mechanical Engineering}

\date{September 2015;
submitted to J. Rheol.}

\begin{abstract}
Recent Couette-cell shear experiments of carbopol gels have revealed the formation of a transient shear band before reaching the steady state, which is characterized by homogeneous flow. This shear band is observed in the small-gap limit where the shear stress is spatially uniform. An effective-temperature model of the transient shear banding and solid-fluid transition is developed for the small-gap limit. The small-gap model demonstrates the ability of a continuum-constitutive law that is based solely on microstructural rearrangements of the gel to account for this transient behavior, and identifies that it proceeds via two distinct processes. A shear band nucleates and gradually broadens via disordering at the interface of the band. Simultaneously, spatially homogeneous fluidization is induced outside of the shear band where the disorder of the gel grows uniformly. Experimental data are used to determine the physical parameters of the theory, and direct, quantitative comparison is made to measurements of the structural evolution of the gel, its fluidization time, and its mechanical response under plastic flow.
\end{abstract}

\maketitle

\section{Introduction}
Yield stress fluids (YSFs) are ubiquitous in everyday life, and their special properties have merited intense research \cite{bonn,rag,moller_phil,moller_soft,coussot,barnes,ovarlez,schall,bird}. 
Examples include gels, clay suspensions, foams, concentrated emulsions, and colloids. These seemingly distinct substances exhibit a similar mechanical response when subjected to shear deformation: Below a critical (yield) stress these materials remain elastic, behaving as solids, but above this critical stress they are able to deform and flow as viscous liquids. This characteristic ability makes them extremely sought-after for many applications \cite{coussot_book}. In the presence of an applied shear protocol a YSF can exhibit a distinctive mode of deformation known as \textit{shear banding}, where a region of highly localized strain that is far greater than that of the surrounding material appears. The region, or band, is often long or continuous in the direction of shear and of some finite width perpendicular to the shear. This form of shear localization has been shown to be a common occurrence in the rheology of complex fluids as well as glassy materials in general \cite{berthier}. 

YSFs are often categorized by the degree of thixotropy they present. Thixotropic YSFs have memory and aging effects that lead to a history dependence, marked e.g. by flow curves that are significantly different when the applied shear rate is ramped up compared to the ramp down. In contrast, nonthixotropic or ``simple'' YSFs have no apparent history dependence over the timescales of observation. They also tend to have predominantly repulsive interactions between their constituent mesoscopic substructures \cite{fielding3}. More important is the fact that thixotropic YSFs, regardless of the particular material, easily form shear bands. Some known as viscosity-bifurcating YSFs, display shear banding in the steady state, a phenomenon which as been observed in a number of complex fluids including polymers \cite{boukany1,boukany2,ofech,tapadia} and wormlike micelles \cite{helgeson,hu,decruppe,britton}. Experiments of simple YSFs on the contrary, report conflicting accounts of the formation of shear bands: In some instances shear bands are observed and yet in others the system (sometimes the same system) remains completely homogeneous under shear \cite{gilbreth,krishan,rodts,ovarlez_mri,nonnewtonian,divoux,divoux2,divoux3,divoux4}. These contradictory findings have prompted investigations of simple YSFs with greater attention to the specific material and composition, the shear procedure, and the influence of the geometry of the apparatuses.    

Carbopol gel is widely regarded as a quintessential simple YSF and as such has generally been believed to transition from the solid to the liquid state uniformly as a homogeneous system \cite{moller_phil,rag}. Recent Couette-cell experiments of shear in a carbopol gel have instead revealed the conspicuous formation of transient shear bands before the gel reaches a uniform steady state characterized by a linear velocity profile and Herschel-Bulkley rheology \cite{divoux}. More importantly because of the small width of the particular Couette cell, the shear stress measured in these experiments was observed to be nearly uniform across the gap, and the stress gradient was reported to not play a significant role in the formation of the transient shear band. 

The onset of plastic flow in YSFs and other similar materials has been investigated by other theoretical approaches, but direct, quantitative comparison has been lacking with regard to shear banding and the process of fluidization. Among these theoretical frameworks are soft glassy rheology (SGR) \cite{fielding1}. The SGR model in particular has been successful in qualitatively describing the general steady-state power-law behaviors seen in experiments, and criteria have been proposed to characterize the onset of shear banding \cite{fielding2}. It has also been able to capture aging. One drawback of SGR is that it assumes the existence of a ``noise temperature" that controls activation rates of plastic processes. The physical basis for this noise temperature is however not clear, and its dynamics have not been derived from fundamental or mesoscale principles. 

A minimal theoretical model of a Newtonian fluid using the Krieger-Dougherty constitutive relation \cite{illa} has been used to qualitatively model transient shear banding in a Couette-cell geometry. The shortcomings of this approach are that it does not describe a yield stress or solid-state elastic response, nor does it quantitatively capture key experimental observations of shear banding in YSFs, i.e. the correct spatial and temporal evolution of the shear band and fluidization times of the system. More importantly, this model attributes transient shear banding to the nature of the Couette-cell apparatus itself. Specifically it suggests that the gradient in the stress field, which exists simply due to a cylindrical geometry, leads to a higher shear rate and unjamming of material near the inner cylinder of the Couette cell. At the present time however, it is unclear to us whether the primary mechanism by which such transient shear banding occurs is the stress gradient caused by the cylindrical geometry, or if the mechanism is an instability in the microstructure of the gel itself. In the case of hard-sphere colloidal glasses similar questions have been raised and subsequent analyses have been preformed to address the possibility that an intrinsic instability via shear-concentration coupling can lead to shear banding in the absence of gradient-driving nonuniform stress fields \cite{bessing_isa}.   

Motivated by the possibility that heterogeneities in the microstructure of the gel may indeed be the dominant driver of the observed shear band, we propose an alternative description of the phenomenon of transient shear banding during plastic flow based on fluctuations in the structure of the gel. We adapt the effective-temperature hypothesis of the shear transformation zone (STZ) theory as a rheological model for shear localization in a simple YSF. In this description, structural changes described by an effective temperature \cite{bouchbinder3} account for the observed plastic flow, and the formation of a transient shear band is the primary mode of deformation of the carbopol gel. This occurs simultaneously alongside a distinct, uniform fluidization of the gel outside the shear band. We then make quantitative comparisons of stress-time behavior and velocity profiles with recent small-gap Couette-cell experiments of a carbopol gel. 

The STZ theory is a general framework for characterizing plasticity in amorphous materials, and provides a continuum-level mean-field approximation for flow based on an assumption of local rearrangements of a material's structure via the activation of STZs \cite{falk2,pechenik,pechenik2,bouchbinder,bouchbinder2,bouchbinder3,bouchbinder4,bouchbinder5,langer,falk_conds,lisa,lisa2,lisa3}. STZs are orientational point defects that mediate plastic flow by accommodating rearrangement. In this paper we use data from the aforementioned experiments in \cite{divoux} to determine the physical parameters of the theory and make precise experimental connections. A distinguishing feature of the STZ theory is that it is based on a specific model of molecular rearrangements, which have been observed directly in numerical simulations and analog experiments \cite{falk_conds,dennin,daub}. The present formulation of the STZ theory has been extended from the original work of Argon \cite{argon,argon2} and his proposal of ``shear transformations" to explain plastic deformation in metallic glasses, as well as from the free-volume and flow-defect theories of Turnbull, Cohen, Spaepen, and others \cite{turnbull,spaepen}. The STZs themselves have been postulated to have internal degrees-of-freedom \cite{falk_conds}. The STZs not only transform from one orientation to another; they are created and annihilated during configurational fluctuations at a rate that is proportional to the rate of energy-dissipation per STZ \cite{pechenik}. A coarse-grained continuum description of the effect of the changing population density and spatial distribution of the STZs is defined by an effective-temperature field. In this paper we present equations of motion for the effective temperature and plastic-strain where the steady state is set by a Herschel-Bulkley rheology, and no explicit dependence of the STZ internal variables, local strain rate, or flow history appears.   

A flow rule for the plastic component of the rate-of-deformation tensor ${\bf{D}}^{{\rm pl}}$, which we will subsequently call the plastic-strain rate as a matter of convention follows from the STZ dynamics. For a monotonically loaded, athermal system where there are no rate-dependent processes such as aging, which compete with the STZ-transition rates, and where we assume there to be a low STZ density, the flow rule can take the form,
\begin{equation}\label{strain_full}
\bf{D}^{\rm pl} = \bf{F} \it{e^{-\rm{1}/\chi}} \ ,
\end{equation}
where $\bf{F}=\bf{F}(\bf{S})$ is a monotonic tensor-function of the deviatoric Cauchy stress $\bf{S}$. One critical way the STZ theory differs from its predecessors (and indeed other theories) is through the introduction of the quantity $\chi$ and its relationship to an ``effective temperature" $T_{\rm eff}$ that is defined as 
\begin{equation}
T_{\rm eff} = \frac{\partial U_{c}}{\partial S_{c}} \ ,
\end{equation}
where $U_{\rm c}$ and $S_{\rm c}$ are the material's potential energy and 
entropy respectively of only the configurational degrees-of-freedom, i.e. those degrees-of-freedom associated with the structure of the material. This is to be distinguished from the usual thermalized temperature $T$ which accounts for the degrees-of-freedom which relax on timescales short compared to the observation time. The typical definition of $T$ is applicable to the fast, i.e. vibrational degrees-of-freedom, but the configurational degrees-of-freedom are typically out of equilibrium. Some limited attempts to experimentally measure an effective temperature for disordered materials have been made \cite{dieterich}, as well as other direct, quantitative comparisons with experiments of bulk metallic glasses \cite{daub}. Atomistic simulations of two-dimensional glasses have already suggested that $T_{\rm eff}$ is linearly related to the local potential energy per atom \cite{falk3}, and earlier many arguments were made for the existence of the notion of an effective temperature, related to the entropy, that characterizes the number of phase-space paths available to the system \cite{lema,langer,edwards_gri,ono}.      

In the effective-temperature STZ formalism, the dimensionless scalar field $\chi$ is defined as $\chi = k_{B} T_{\rm eff} / E_{z}$, where  $k_{B}$ is the Boltzmann factor. Although $\chi$ is a dimensionless form of $T_{\rm eff}$ we shall subsequently refer to it as simply ``the effective temperature'' for readability.  Here $E_{z}$ is a typical energy required to create an STZ. In the athermal limit the dynamical equation for the effective temperature $\chi$ takes the form
\begin{equation}\label{chi_full}
\dot{\chi} =\frac{\bf{S} : \bf{D}^{{\rm pl}}}{c_{\rm eff}} \left(\chi_{\infty} - \chi \right)  + D_{\chi} \nabla^{2} \chi \ .
\end{equation}  
The first term on the RHS in Eq.~\ref{chi_full} represents a source of plastic work per unit time that does mechanical work on the structural degrees-of-freedom. The parameter $c_{\rm eff}$ is the volumetric effective-heat capacity with dimensions of energy per unit volume, determining the energy input per unit increment of effective temperature. In flowing regions $\chi$ converges to a limiting value $\chi_{\infty}$, which represents the steady-state effective temperature where the work done to shear the structure no longer causes an increase in disorder. In developing the analysis that follows we have assumed for simplicity that $\chi_{\infty}$ is a constant for the entire system. In general $\chi_{\infty}$ is usually rate dependent \cite{lisa2}, however in the low-rate limit $\chi_{\infty}$ is approximately a constant. This assumption is consistent with the lower-limiting value of effective temperature found for simulations of sheared glasses in the steady state \cite{liu}. The initial value $\chi_{o}$ characterizes the structure of the gel in the pre-sheared state, and ideally would come from an analysis of the microstructural information of the system's constituents. In the absence of this molecular-level information, the form of $\chi_{o}$ including the mean value and any seeding of fluctuations about the mean are usually chosen in a way to best match the macroscopic behavior, e.g. the stress-strain curves of the material \cite{lisa,lisa2,lisa3,daub}. Although the range of values of $\chi_{o}$ for a particular system is significantly restricted by the nonlinear form of Eq. \ref{chi_full} and its stability. 

The final term in Eq.~\ref{chi_full} describes the diffusion of the effective temperature through an effective thermal diffusivity $D_{\chi}=k_{\rm eff}/c_{\rm eff}$ with dimensions length-squared per unit time, where $k_{\rm eff}$ is the effective thermal conductivity. Here we consider the simplified case where $ D_{\chi}$ is a constant and diffusion occurs only in the presence of changes in the gradient of $\chi$. The diffusivity $D_{\chi}$ itself sets a lengthscale for the simulation that is approximately the size of an STZ. We plan to address the general case of the possibility of a strain-rate-dependent diffusivity in a subsequent paper.  

The effective-temperature theory as a rheological model is established in Sec.~\ref{basic_theory}. Section~\ref{model_simulations} presents the results of the small-gap effective-temperature model, making direct and quantitative comparisons with the recent experimental results of \cite{divoux}, including stress, fluidization time, and velocity data. The unique two-state fluidization that occurs during the transient is described and explained in terms of the dynamics of the effective-temperature constitutive law. The robustness of the model, as parametrized for a particular applied shear rate, is discussed in terms of the power-law behavior found by the experimenters.  We conclude in 
Sec.\ref{conclusions} with a discussion of the model and its implications for understanding transients during plastic flow in YSFs, and comment on the on-going development of this model.

\section{Basic Theory}
\label{basic_theory} 
\begin{figure}
\begin{center}
\includegraphics[scale=0.60]{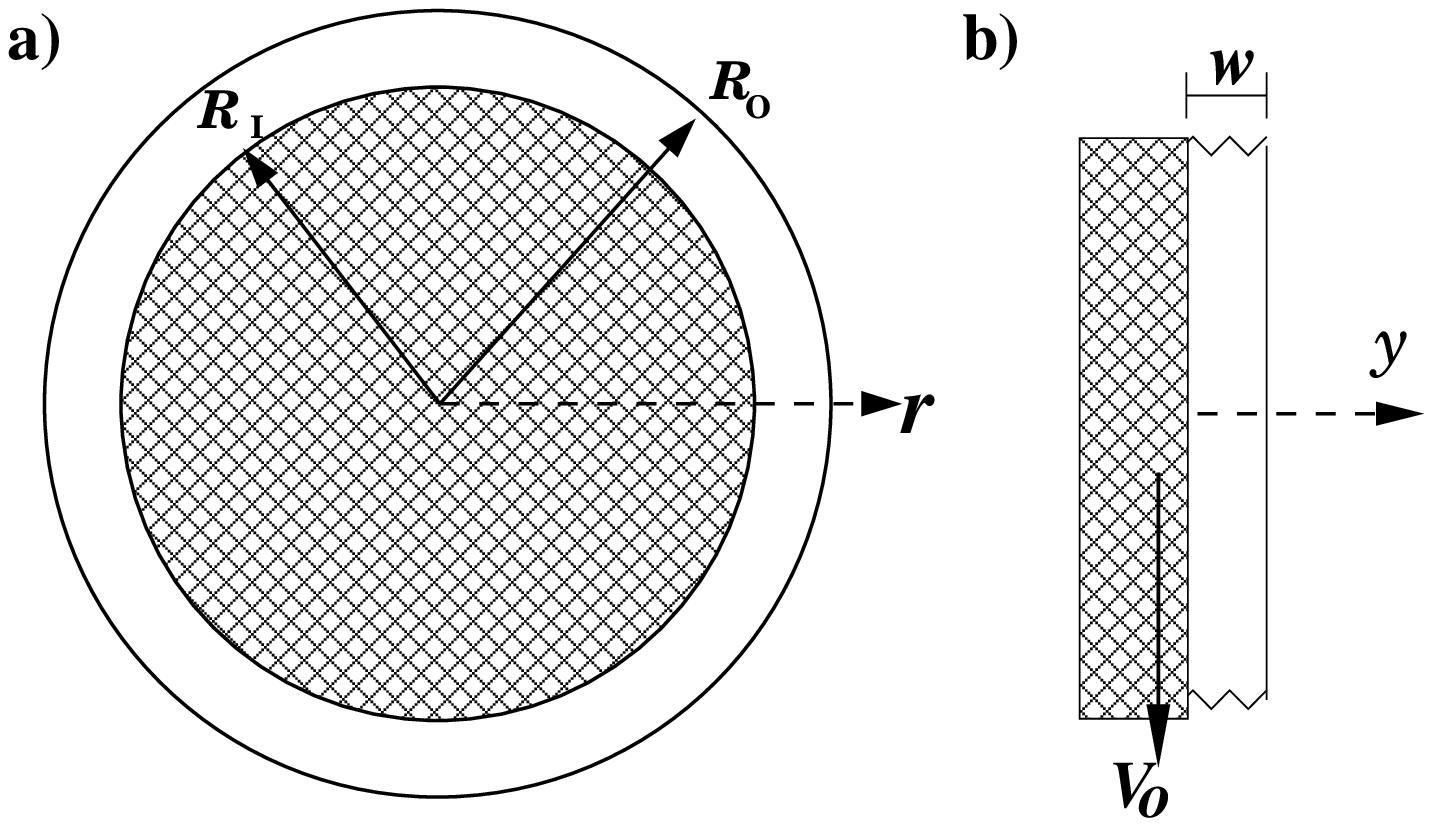}
\caption{\small{a. The edge of the inner Couette-cell cylinder ($R_{\rm I}=23.9$ mm) and outer cylinder edge ($R_{\rm o}$) with gel (white) where the gap width $w = R_{\rm o} - R_{\rm I}$ = 1.1 mm. The inner cylinder is held at constant velocity $v_{o} = 0.77$ mm$/$s. The nucleation of a shear band begins at $R_{\rm I}$ and then broadens outward in the $y$-direction. b. The Couette-cell system in the small-gap limit where the curvilinearity vanishes, transforming the Couette-cell problem to a simple shear geometry. Here the system is fixed at $y=w$ and driven with speed $v_{o}$ at the boundary $y=0$.}}
\label{mycouette}  
\end{center}
\end{figure}

In this section we adapt the effective-temperature definition taken from the STZ theory for a small-gap Couette-cell system, whose geometry is described in Fig.~\ref{mycouette}a. The symmetry of the Couette cell allows us to treat the problem as an effectively one-dimensional system. We deliberately move to the small-gap limit where the curvilinearity vanishes as shown in Fig.~\ref{mycouette}b. This is consistent with experimental findings of a negligible inhomogeneity in the stress field across the gap such that the stress was reported as being uniform \cite{divoux,divoux2,divoux3,divoux4}. The complete absence of any stress gradient in the model allows us to test whether the transient shear banding can be accounted for purely by a microstructural heterogeneity occurring near the moving rotor. Indeed, we find that the effective-temperature theory is able to reproduce the experimental phenomenon in nearly every detail without including the stress gradient. We further assume that we are in the inertial limit, namely $\nabla \cdot \bf{S}=0$ which is also consistent with the experimental conditions. 

A key experimental observation of the rheology of simple YSFs is that the steady-state behavior of the system is described by the well-known Herschel-Bulkley (HB) relation \cite{divoux,divoux2,divoux3,divoux4}, and this will be used to determine a flow rule for $\bf{D}^{{\rm pl}}$.     
  
\subsection{Plastic-Strain Rate}
The one-dimensional plastic-strain rate $\dot{\epsilon}^{{\rm pl}}$ from the STZ theory 
takes the form 
\begin{equation}\label{strain}
\dot{\epsilon}^{{\rm pl}} =  f(s) e^{-1/\chi} \ .
\end{equation}
We next assume that the steady-state behavior of the gel is given by a HB form, namely $s = s_{c} + 
A\left(\dot{\epsilon}^{{\rm pl}}\right)^{n}$ for some parameters $n$ and $A$ 
which characterize a system that begins to unjam and flow plastically 
when the deviatoric stress $s$ reaches $s_{c}$, the critical stress.  Here the HB coefficient $A$ has units of Pa$\cdot$s$^{n}$, and sets the timescale for the fluidization of the gel. The plastic-strain rate is related to the measure of 
disordering by a function $f=f(s)$. We 
require that Eq.~\ref{strain}, a form of the constitutive law in the STZ theory, 
reduce to a HB relation when $\chi = \chi_{\infty}$. Therefore $\dot{\epsilon}^{{\rm pl}}$ 
in the steady state 
$\dot{\epsilon}^{{\rm pl}}_{\infty}$ must be

\begin{equation}
\dot{\epsilon}^{{\rm pl}}_{\infty} = \left(\frac{s-s_{c}}{A}\right)^{1/n} \ .
\end{equation}
This allows $f$ to be determined so that 
\begin{equation}
 \dot{\epsilon}^{{\rm pl}}_{\infty} =  f(s) e^{-1/\chi_{\infty}} = \left(\frac{s-s_{c}}{A}\right)^{1/n}
\end{equation}
and we find 
\begin{equation}\label{f}
f(s) = \left(\frac{s-s_{c}}{A}\right)^{1/n} e^{1/\chi_{\infty}} \ .
\end{equation}
Upon substituting the expression for $f$ into Eq.~\ref{strain}, we arrive at the  expression for the plastic-strain rate, namely
\begin{equation}\label{strain_final}
\dot{\epsilon}^{{\rm pl}} = \left(\frac{s-s_{c}}{A}\right)^{1/n}e^{1/\chi_{\infty}-1/\chi} \ .
\end{equation}
This result provides a continuum constitutive-law relating the plastic-strain rate to the deviatoric stress within the framework of an effective temperature.  

\subsection{Stress Rate}
To derive the differential equation for the deviatoric-stress rate we express the strain rate as the sum of the elastic and plastic components, 
namely
\begin{equation}\label{totstrain}
\dot{\epsilon} = \dot{\epsilon}^{{\rm el}} + \dot{\epsilon}^{{\rm pl}} \ .
\end{equation}
The velocity $v=v(y)$ as a function along the Couette-cell radius $y$ in 
Fig.~\ref{mycouette} is then given by 
\begin{equation}
v(y) = \int \dot{\epsilon} \ dy = \int \dot{\epsilon}^{{\rm el}} \ dy + 
\int \dot{\epsilon}^{{\rm pl}} \ dy 
\end{equation}
and when integrated across the full gap-width, one finds
\begin{equation}
v_{\rm o}= \frac{\dot{s} w}{\mu} + \left(\frac{s-s_{c}}{A}\right)^{1/n} 
e^{1/\chi_{\infty}} \int_{R_{\rm o}}^{R_{\rm I}} e^{-1/\chi} \ dy
\end{equation}
where $\mu$ is the shear modulus (with dimensions of 
stress), and $\bar{\dot{\gamma}} = v_{o}/w$ is the average strain-rate across the gap found by imposing a velocity $v_{o}$ at the inner cylinder. The shear experiment reported in \cite{divoux}, to which we are seeking to compare in detail, uses a  $\bar{\dot{\gamma}}=0.7$ s$^{-1}$. We use this rate in all of our simulation results, except for the stress-time profiles of Fig~\ref{powerlaw} where each curve corresponds to a unique $\bar{\dot{\gamma}}$. Because plastic flow is the dominant flow process for the transient shear band formation reported in \cite{divoux}, we have assumed for simplicity that the elastic response is linear and that all non-linear behavior results from the plastic response. While the ramp-up of the elastic response is very brief, linear-elastic strains persist throughout the simulation and act to drive the plastic flow. The shear banding itself is an inherently plastic phenomenon and the relevant timescale for the transient dynamics comes from the plastic response. Rearranging 
we find that 
\begin{equation}\label{stress}
\dot{s} =  \mu \bar{\dot{\gamma}} - \mu \left(\frac{s-s_{c}}{A}\right)^{1/n} 
e^{1/\chi_{\infty}} \frac{1}{w} \int_{R_{\rm o}}^{R_{\rm I}} e^{-1/\chi} \ dy \ .
\end{equation}

\section{Model Simulations}
\label{model_simulations}
The equations of motion (EOM) of the effective-temperature model that we have presented in 
Sec.~\ref{basic_theory} 
are
\begin{equation}
\label{strainf}
\dot{\epsilon}^{{\rm pl}} = \left(\frac{s-s_{c}}{A}\right)^{1/n}e^{1/\chi_{\infty}}e^{-1/\chi}
\end{equation}
\begin{equation}\label{stressf}
\dot{s} =  \mu \bar{\dot{\gamma}} - \mu \left(\frac{s-s_{c}}{A}\right)^{1/n} 
e^{1/\chi_{\infty}} \frac{1}{w} \int_{0}^{w} e^{-1/\chi} \ dy
\end{equation}
\begin{equation}\label{chif}
\dot{\chi} = \frac{2 s}{c_{\rm eff}} \left(\frac{s-s_{c}}{A}\right)^{1/n} e^{1/\chi_{\infty}-1/\chi}\left(\chi_{\infty} - \chi \right) + D_{\chi} 
\frac{\partial^{2}\chi}{\partial y^{2}} \ .
\end{equation}

\begin{table}
\begin{center}
\scalebox{0.8}{
    \begin{tabular}{|c|c|c|c|}
           \hline\hline
        PARAMETERS & SYMBOL & UNIT & VALUE \\ \hline    
        Herschel-Bulkley exponent   & $n$ & -  & 0.55   \\ 
        Herschel-Bulkley yield stress     & $s_{c}$&Pa     & 29.4  \\ 
        Herschel-Bulkley coefficient      & $A$&Pa$\cdot$s$^{n}$  & 1.60  \\ 
        Diffusivity  & $D_{\chi}$  &mm$^{2}$s$^{-1}$     & $10^{-6}$  \\  
        Shear modulus       & $\mu$   &Pa       & 95 \\
        Volumetric effective-heat capacity    & $c_{\rm eff}$ &Pa  & 5 $\times 10^{3}$   \\ 
        Steady state $\chi$  & $\chi_{\infty}$& - & 0.064  \\         
        \hline\hline
    \end{tabular}
    }
    \caption{The parameters characterizing the small-gap effective-temperature model's constitutive law are presented with the values found to best match the stress-time and velocity measurements of the experiments. The ``-" denotes the parameter is dimensionless. $A$, $n$, and $s_{c}$ are from the Herschel-Bulkley rheology. The $c_{\rm eff}$, $D_{\chi}$, and $\chi_{\infty}$ are from the effective-temperature constitutive model.} 
    \label{parameters}
    \end{center}
\end{table}

The three EOM take the form of coupled, nonlinear, partial integro-differential equations which we integrate in time. At time $t=0$, $s=0$ so that the fluid is initially unstressed in our analysis. Boundary conditions  for $\chi$ must also be placed at the rotor and the outer wall of the Couette cell. We impose a no-conduction boundary condition $\partial \chi / \partial y =0$, since $\chi$ is restricted to the gel itself.   

\begin{figure}
\begin{center}
\includegraphics[scale=0.6]{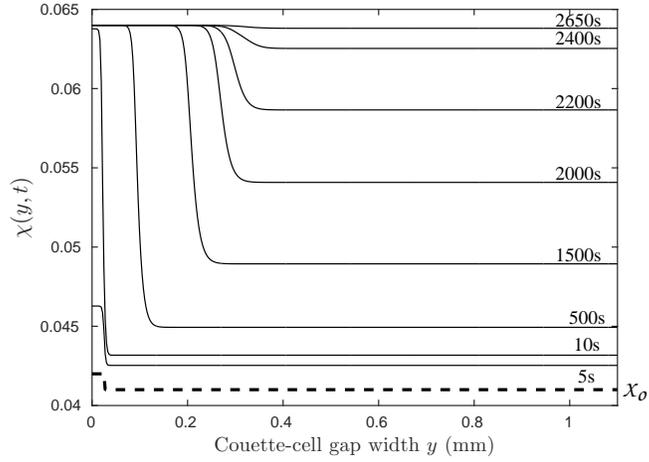}
\caption{\small{The effective temperature $\chi$ at various times during the shear-induced deformation of the small-gap model simulation. The initial field $\chi(y,0)=\chi_{o}$ (heavy black dashed line) containing a perturbation near the inner wall ($y=0$) into a shear band as time evolves. The increasing value of $\chi$ shows the band broadening across the gap ($y$-direction) before reaching a uniform steady-state. A strain rate $\bar{\dot{\gamma}}= {\rm 0.7 \ s^{-1}}$ is applied at the rotor.}}
\label{chi_fig}  
\end{center}
\end{figure}

As discussed in the Introduction the initial value of the effective temperature $\chi_{o}$ should ideally follow from a detailed analysis of the microstructural information about the gel. Without such detail, we are left to conjecture an appropriate $\chi_{o}$. We began by following the work of Manning et al. \cite{lisa}, where we assume the initial condition can be decomposed into a background term and a fluctuation term about the background
\begin{equation}\label{decompose}
\chi_{o}= \chi^{\rm BG}_{o} + \delta \chi_{o}  \ \ .
\end{equation}
To simulate the disordering of the gel near the inner cylinder (rotor) we chose the simplest type of fluctuation term---a perturbation near the inner cylinder. Equation ~\ref{decompose}, which is shown in Fig.~\ref{chi_fig} was given the explicit form of a spatially uniform background $\chi^{\rm BG}_{o}=0.041, \ \forall y: \ 0 \leq y \leq w$ with a perturbation $\delta \chi_{o}=0.001, \ \forall y: \ 0 \leq y \leq 0.03$mm, near the inner cylinder where disordering of the structure would likely first occur, and thus nucleate a shear band \cite{divoux}. We found that this value for $\chi^{\rm BG}_{o}$ generated a stress-time curve that reasonably matched the range from the experiment, and the value of $\delta \chi_{o}$ was as small as possible without inducing purely homogeneous fluidization. The surfaces of both the inner and outer cylinders also present sites for structural heterogeneities. The slightly larger stress at the rotating inner cylinder favors localization there. Indeed, a boundary layer is known to form at the inner cylinder in the experiments, although we do not explicitly model the layer here. As discussed earlier, we ignore the stress gradient in our numerical analysis as it is assumed to be negligible for a small-gap width, other than to preference the nucleation at the inner cylinder. The ability of this fluctuation at the inner rotor to grow and lead to strain localization in the form of a shear band depends on both the size of the perturbation and the average value of $\chi_{o}$, in addition to the chosen constitutive law.  It was shown in detail that small-amplitude perturbations can possibly grow during the transient regime if $\chi_{o}$ and its fluctuations meet an explicit criterion \cite{lisa}. However, even if they grow, implying that the equation of motion for $\chi$ is unstable over some time interval, it does not guarantee that the fluctuations will form a shear band. According to \cite{lisa}, the magnitude of $\chi_{o}$ must be below a limiting value for linear stability during the transient regime in order for strain localization to be possible. Furthermore the $\chi_{o}$ chosen here has a numerical localization number which is within the range corresponding to at least partial to full strain-localization. We found this $\chi_{o}$ field matches experimental width measurements of the shear band at early times, while also keeping the perturbation at the inner wall small enough to be a linear instability. Since larger values of the perturbation only enhance localization, we have deliberately chosen a value that would conceivably test the growth of a linear instability into a fully developed shear band. Alternatively a universal criterion for shear band formation in time-dependent fluids has been proposed by \cite{fielding2} whereby the shape of the stress-time curve determines when shear banding can initiate. In particular the condition that $\partial_{\gamma} s + \dot{\gamma} \partial^{2}_{\gamma} s < 0$ was calculated to hold $\forall t:  t \geq 3$ s. The first noticeable growth in $\chi$ (but not yet reaching $\chi_{\infty}$) occurs at 5 s and is shown in Fig.~\ref{chi_fig}, exactly after the linearly elastic regime ends and flow is initiating in the form of a shear band. This criterion is therefore consistent with the onset of shear banding in the effective-temperature model.

\begin{figure}[!h]
\begin{center}
\includegraphics[scale=0.6]{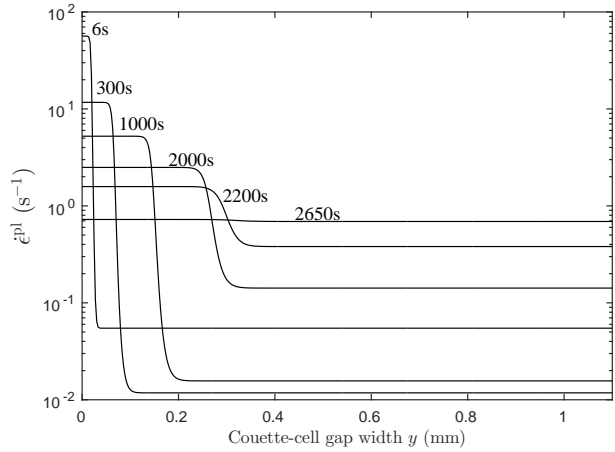}
\caption{\small{Plastic-strain rates at different times across the gap during the small-gap model simulation. As plastic strain develops in the gel, $\dot{\epsilon}^{\rm pl}$ takes on a large value near the shearing rotor ($y=0$). As a shear band forms and traverses the gap the system fully fluidizes via a two-state process, until steady state is reached and everywhere $\dot{\epsilon}^{\rm pl}=0.7$, the value of the applied shear-rate.}}
\label{strain_few}  
\end{center}
\end{figure}

Table~\ref{parameters} contains the values of the parameters which appear in the small-gap model's EOM. It includes those parameters which arise from the effective-temperature phenomenology of the model $c_{\rm eff}$ and $D_{\chi}$, as well as those associated with a HB Law $A, n$, and $s_{\rm c}$. Once the initial effective temperature field $\chi_{o}$ was chosen and found to produce results which were within the range of the experimental stress-time profile, the parameters of the model were fit to the experimental velocity profiles and fluidization time, and to the precise inflection and shape of the stress-time curve. The central parameter in the effective-temperature hypothesis is the volumetric effective-heat capacity $c_{\rm eff}$ which appears in the equation for $\dot{\chi}$. The quantity $c_{\rm eff}$ has the physical significance of being the amount of plastic work per unit volume required to cause a fractional relaxation of $\chi$ to its steady state, and its value was found to significantly determine the fluidization time and the abrupt transition of the entire system to the steady state. The shear modulus $\mu$ was set to match the experimental stress-time curve.

\begin{figure}
\begin{center}
\includegraphics[scale=0.6]{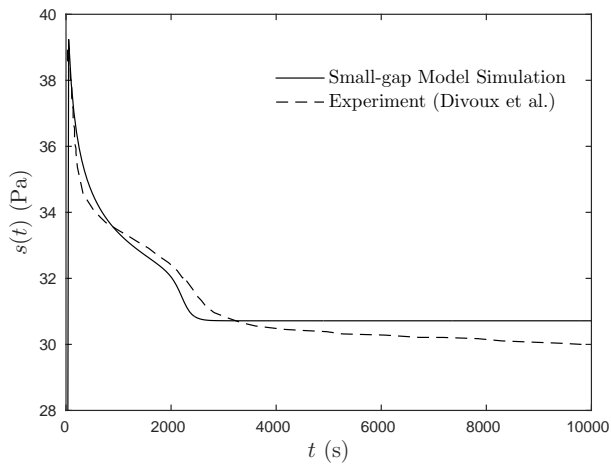}
\caption{\small{The deviatoric shear stress as a 
function of time for the small-gap model and the experimental stress measurements \cite{divoux} under an applied strain-rate $\bar{\dot{\gamma}}= 
{\rm 0.7 \ s^{-1}}$ at the rotor. Complete fluidization of the gel is experimentally observed between 2500-3000s, and is well captured by the simulation.}}
\label{stressfig}  
\end{center}
\end{figure}

The experimenters also reported a flow curve in a separate set of experiments from which they extracted values for a HB steady-state rheology. The values of these experimentally reported parameters, $A=9.8$ (Pa s$^{n}$), $n=0.55,$ and $s_{\rm c}=26.9$ (Pa) are inconsistent with the experimentally reported ``steady-state" condition in the shear experiment that was used to produce the stress-time curve in Fig.~\ref{stressfig}. This is in part due to a waiting time of 100s before each stress measurement was made in the flow curve experiment, compared to the shear experiment, which we are modeling here, where the steady state does not emerge until 10,000s. The values of $A$ and $s_{c}$ are also likely dependent upon the particular sample and its preparation while the exponent $n$ is believed to be more robust. Our constitutive law for the plastic-strain rate assumes some HB form must also hold for the longer-time protocol in the shear experiment, and hence cannot account for this discrepancy between the two types of experiments. As a starting point we initially used the HB value for $A$ reported from the flow-curve experiments and then adjusted it to match the long-time behavior of the experiment plotted in Fig.~\ref{stressfig}, from which the value of $s_{c}$ was extracted. We also confirm the robustness of the exponent $n$, as the same value of $n=0.55$ reported experimentally also best matched the data in the model.

In the shear experiment where the rotor's velocity was held at $v=0.77$ mm/s, the gel was initially prepared to remove memory effects using the same protocol before each experiment \cite{divoux}: A pre-shear lasting 60 s at 10$^{3}$ s$^{-1}$ was applied in the clockwise direction. Then 60 s at 10$^{3}$ s$^{-1}$ was applied in the counterclockwise direction. The shear was then instantaneously stopped and the gel was allowed to rebuild for 120 s. The experiments then involved shearing at a constant rate for 10$^{4}$ s while the stress was measured. A shear band was observed to nucleate near the rotating inner wall of the cell. After nucleation the band widened until the onset of fluidization, at which point the entire gel transitioned to a homogeneously flowing state, revealing a distinct process of unjamming. During the transient characterized by the broadening of the shear band, the shear rate and velocity profile of the gel outside the band were found to be non-zero. 

The gel's restructuring under shear is captured by the evolution of $\chi$ during the numerical simulations. In Fig.~\ref{chi_fig}, $\chi$ is plotted at different times during the deformation. The bottom-most, dashed black curve in Fig.~\ref{chi_fig} illustrates $\chi(y,0)=\chi_{o}$, the state before shearing begins. The time-evolution of the plastic-strain rate is described in Fig.~\ref{strain_few}, and reflects these changes in $\chi$. Central to the theory is the notion that the local plastic-strain rate depends on the changing state of the amorphous structure, which is described by $\chi=\chi(y,t)$ . At any time, $\chi$ quantifies the disordering of the material as it is strained or stirred. In the athermal limit, the time-evolution $\dot{\chi}$ will be nonzero only if a state of stress greater than a critical stress is present or if there are spatial variations large enough to activate diffusion through $\nabla^{2} \chi$. Stresses above the critical stress induce microstructural rearrangements which cause $\chi$ to evolve towards its steady-state value $\chi_{\infty}$.    

In our effective-temperature model the steady state is determined by the two terms on the RHS of Eq.~\ref{chi_full}, a source term $\bf{S}:\bf{D}^{{\rm pl}} \left(\chi_{\infty} - \chi \right)/{\it c}_{\rm eff}$ and a Laplacian term $D_{\chi} \nabla^{2}\chi$ which acts to diminish gradients in $\chi$. As such, the steady state simply occurs when $\chi=\chi_{\infty}$, since the Laplacian of $\chi$ becomes negligible in the steady state.

As was observed experimentally, the simulations reveal the formation of a shear band, which nucleates from the perturbation near the inner cylinder and begins to grow, broadening across the gap. In the simulations this is followed by an abrupt transition during which the gel reaches a fully unjammed state and completely fluidizes. We found the value of $\chi_{\infty}=0.064$ to best match the sudden transition to homogeneous flow. The stress-time curve of the simulation also agrees with the experimental measurements as seen in Fig.~\ref{stressfig}, where in the model the gel undergoes an extremely brief, linearly elastic regime followed by plastic flow (including the transient shear banding) until the steady-state stress given by the HB relation is reached. The time needed for the shear band to grow sufficiently to accommodate the strain rate imposed across the gel results in the overshoot in the stress that is seen in Fig~\ref{stressfig}. This is related to the value of the initial condition for the effective temperature. For example, a stiffer, stronger, more ordered system (lower $\chi_{o}$), will have a larger stress-overshoot.

\begin{figure}
\begin{center}
\includegraphics[scale=0.6]{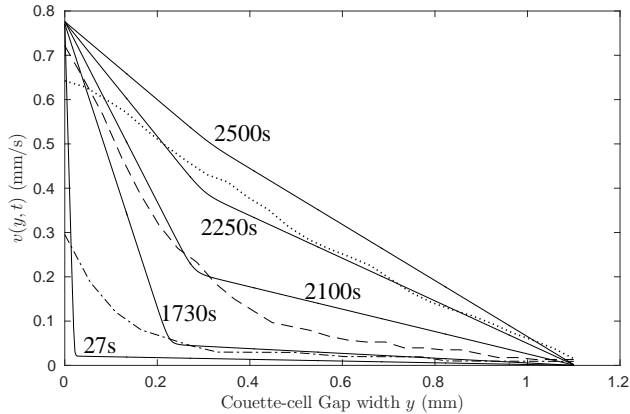}
\caption{\small{Velocity profiles at different times during the small-gap model simulation (solid curves) compared to experimental measurements at times 27 s ($-\cdot-$), 1,730 s ($--$),  and 2,250 s ($\cdot\cdot\cdot$). The velocity evolves from a discontinuous profile over the gap, indicating the coexistence of a shear band with more gradual fluidization in the material outside the band. There is significant wall slip at early times in the experiments. A nearly linear profile in the steady state is observed in both the experimental measurements and the simulations.}}
\label{velocity_all}  
\end{center}
\end{figure}

The sudden fluidization of the gel after the formation of the shear band is also evident in the velocity profiles shown in Fig.~\ref{velocity_all}. The portion of the velocity curves with the significantly steeper slope near the inner cylinder is indicative of a shear band, while the lower slope near the outer cylinder suggests that the gel outside the shear band is also gradually flowing as it fluidizes. Experimental measurements confirm the non-zero velocity for the gel in front of the shear band. Direct, quantitative comparisons of the velocity profiles of the simulations and the published experimental data are difficult because of the significant wall slip found in the experiment especially at early times, effects which are not included in the model. In Fig.~\ref{velocity_all} we have compared three instances from the simulations against the experimental data with reasonable agreement as the fluidization time is approached, experimentally between 2500-3000 s. 

\begin{figure}
\begin{center}
\includegraphics[scale=0.32]{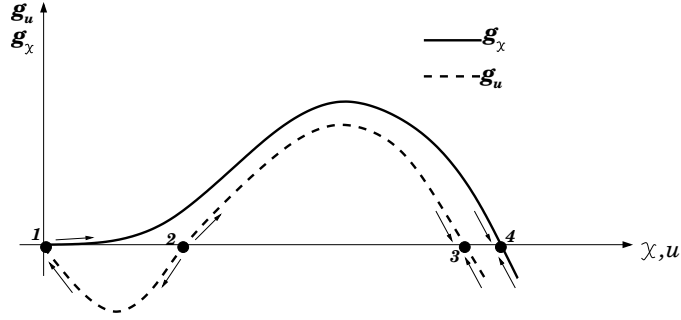}
\caption{\small{Fluidization outside the transient shear band: A comparative sketch of the form of the F-KPP equation $\partial_{t}u-\partial^{2}_{x}u=g_{u}(u)$ (dashed curve) and $\partial_{t}\chi-D_{\chi}\partial^{2}_{x}\chi=g_{\chi}(\chi)$ (solid curve) in the effective-temperature theory. Points 1 and 3 represent stable fixed-points for $g_{u}(u)$ which can describe distinct stable phases of a system into which material that is in a metastable state, represented by the unstable fixed-point at 2, could be transforming. In the effective-temperature model however, only one fixed point exists at 4; it is stable and corresponds to the shear band where $\chi=\chi_{\infty}$, the flowing steady state. For small $\chi$ near point 1, corresponding to the gel outside the shear band, $g_{\chi}(\chi)\rightarrow 0$ due to $e^{-1/\chi}$. However no fixed point exits at 1 for $g_{\chi}(\chi)$, and this feature allows the gel outside the shear band to slowly fluidize through a distinct process of its own.}}
\label{fkpp_fig}  
\end{center}
\end{figure}

We see the presence of the two distinct fluidization processes in the model as a result of the effective-temperature dynamics and the particular form of the constitutive law from the STZ theory that we are using. The equation of motion for $\chi$ takes the form of a second-order parabolic partial differential equation (e.g. the heat equation) with source and Laplacian terms as seen in Eq.~\ref{chi_full} or \ref{chif}. In the case of spatially homogeneous solutions, like those found in the steady state, Eq.~\ref{chif} reduces to $\dot{\chi} = g_{\chi}(\chi)$, where $g_{\chi}(\chi)=\frac{2 s}{c_{\rm eff}} \left(\frac{s-s_{c}}{A}\right)^{1/n} e^{1/\chi_{\infty}-1/\chi}\left(\chi_{\infty} - \chi \right)$ is a nonlinear, inhomogeneous term in $\chi$. Here, the function $g_{\chi}(\chi)$ is similar in form (although asymptotically different) to the logistic reaction term found in the F-KPP (Fisher-Kolmogorov-Petrovsky-Piscounov) equation that describes solidification and reaction-diffusion fronts \cite{fkpp,fkpp2,fkpp3}. In the case of the F-KPP equation, a consideration of the stability of equilibria reveals that there is one unstable fixed-point corresponding to a metastable state, and two stable fixed points into which material can be transforming as illustrated in Fig.~\ref{fkpp_fig}. Similarly, $g_{\chi}$ also has a stable fixed-point which corresponds to the shear-banding region where $\chi$ has reached the steady-state value $\chi_{\infty}$. But instead of a metastable state outside the shear band, $g_{\chi}\rightarrow 0$ exponentially, as $\chi\rightarrow 0$. Consequently the gel outside the shear band is extremely sluggish, although it remains subject to gradual fluidization in the presence of a shear stress, even far from the shear-banding region. Ultimately it is the competition between the slow fluidization of $\chi$ outside the shear band and the forward growth of the shear band that begins at the inner cylinder, that acts as the crucial element in the effective-temperature model leading to the sudden transition to homogeneous flow. 

\begin{figure}
\begin{center}
\includegraphics[scale=0.5]{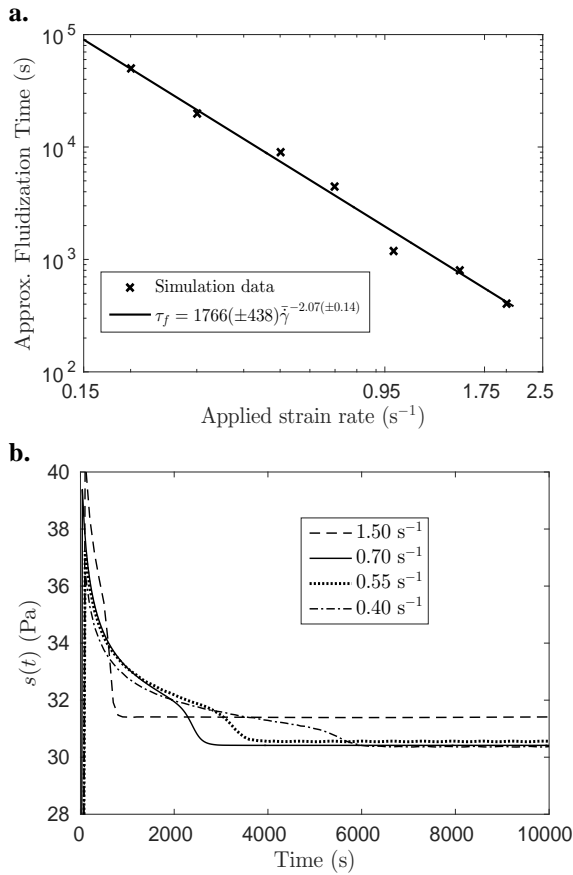}
\caption{\small{ a. The simulated fluidization times $\tau_{f}$ of the model which was parameterized for the case $\bar{\dot{\gamma}}=0.7$ s$^{-1}$ where $w=1.1$mm, and assuming a rough boundary condition. Experimenters have reported a universal power law $\tau_{f}\sim\dot{\gamma}^{-2.3}$ for a wider range of shear rates, various boundary conditions, and several different gap widths. b. The stress-time curves at different imposed strain-rates (inner cylinder velocities) from which the fluidization times (at the onset of the steady state) were extracted.}}
\label{powerlaw}  
\end{center}
\end{figure}

From Fig.~\ref{stressfig} we see the simulation reaches the steady-state stress after approximately 2,600 s. This is also observed from the velocity profile across the gap shown in Fig.~\ref{velocity_all}. The experimentally reported fluidization time for an applied shear-rate of $\bar{\dot{\gamma}}=0.7$ s$^{-1}$ was also approximately between 2500-3000 s. The experiments report a power law for the fluidization time $\tau_{f}\sim\dot{\gamma}^{-2.3}$ as a function of the applied shear rate $\dot{\gamma}$ over a range $0.08 \leq \dot{\gamma} \leq 10$ (s$^{-1}$) using several boundary conditions and sizes of gap width. The effective-temperature model is able to recover approximately the same exponent for a subset of the rates from about $0.1 \leq \bar{\dot{\gamma}} \leq 2.0$ centered on $\bar{\dot{\gamma}}=0.7$, the specific $\bar{\dot{\gamma}}$ for which the model's parameters have been determined with $w=1.1$mm and where a no-slip boundary condition is implicitly assumed. The faster the shear rate in the model is, the faster fluidization and homogeneous flow is reached, and consequently the transient shear band travels a shorter distance across the gap.  Figure~\ref{powerlaw} shows the power-law fit for our simulations.  Outside this subset the $\tau_{f}$ predicted by the model begins to deviate by about an order of magnitude. Interestingly however, the experimental fluidization times outside this subset can still be recovered if $c_{\rm eff}$ is adjusted (re-parametrized) to a lower value for very small shear rates, and increased for the largest shear rates. This would imply that the fraction of plastic work that goes into disordering the gel increases (decreases) as the strain rate increases (decreases), and the power law of the fluidization time $\tau_{f}\sim\dot{\gamma}^{-\alpha}$ may arise in part from this aspect of the physics. This could reflect rate effects in the gel's mechanical response which cause the gel to accommodate deformation through structural disordering at high rates that it could accommodate through independent relaxation processes at low rates. An analysis of the effect of strain-rate dependence is beyond the scope of the theme of this paper, since the model here has not explicit rate-independent. We nonetheless think this is an important direction for future work. To our knowledge no other theoretical models describing the solid-fluid transition make any quantitative predictions for the fluidization times which agree (within an order of magnitude) with these experiments over any range of rates. 

\section{Conclusions}
\label{conclusions}

We have presented a phenomenological effective-temperature model for the small-gap limit of Couette-cell shear experiments of a carbopol gel where transient shear banding is observed. This theoretical model is based on the STZ description, and results in two distinct fluidization processes: the shear band formation and the simultaneously competing homogeneous fluidization. We have made direct, quantitative comparisons with experiments which demonstrate reasonable agreement with the stress-time behavior and velocity profiles of the gel.   

In this theory, consistent with the interpretation of \cite{divoux}, stress gradients are not the primary cause of the strain localization or the fluidization. The small-gap limit of the effective-temperature model strongly suggests that the transient shear banding and sudden fluidization is primarily a result of microstructural disordering originating from structural heterogeneities at the inner cylinder of the Couette cell where the rotor applies the shear. While the gradient is at least negligibly present kinematically, our analysis suggests that it may play little role in the constitutive relation for a YSF under shear---and no role for a small-gap system.  This is in contrast to other theoretical descriptions in which the strain localization is directly associated with the fluidization and comes solely from the non-linear response of the fluid coupled to the stress inhomogeneity that arises from the cylindrical geometry of the Couette-cell \cite{illa}. One possible experimental check would be to see if such transient shear banding and fluidization would take place if the experiment \cite{divoux} were repeated for a different shear geometry, such as shearing the gel between two parallel plates. The absence of any normal stresses from the model's constitutive law suggests to us that such stresses are not necessary to account for the transient shear banding, as was also found by the $\lambda$-based rheological models discussed in the Introduction.

The effective-temperature theory also improves several aspects of existing visco-plastic models which lack any notion of a yield stress, a defining physical feature of YSFs. The mechanical response of the gel is quantitatively well captured by the small-gap simulations, as evidenced by the stress-time curve. The small-gap simulations in this paper reveal the two-stage fluidization seen in experiments, and we have shown that in the model it is directly attributable to the particular form of the constitutive law postulated by the effective-temperature theory. The parameters of the theory provide a physical connection to the thermodynamics of the gel's structural state under shear. One open question is the origin of the strain-rate dependence of the fluidization time for very large and very small rates . A possible explanation is that the volumetric effective-heat capacity, describing the fraction of plastic work that disorders the gel and emerges phenomenologically in the effective-temperature dynamics, is strain-rate dependent and can no longer be approximated as uniform over a wide-enough range of rates.  

The effective-temperature approach that we have proposed offers an important rheological model for both current and future experiments, and responds to a need that was identified by the experimenters for developed theoretical models of transient shear banding in simple YSFs \cite{divoux}. We also believe our approach would benefit from further development, and are currently investigating an extension of this approach for finite-gap systems with non-negligible stress gradients. We are also currently attempting a version of this theory with a rate-dependent diffusivity and steady-state effective temperature.  Beyond this a number of consequential phenomena exist which we have largely ignored in the present model, e.g. aging and wall slip, both of which are known to play at least some role in the recent shear experiments of carbopol. A generalization of this model and incorporation of such additional physics is also a matter of our on-going research.

\section{Acknowledgments}The authors kindly thank Thibaut Divoux and Sebastien Manneville for detailed clarifications about their experiments, as well as M. Lisa Manning and Suzanne Fielding for discussions regarding the nature of the shear-banding instability. This work was supported by NSF Grant Award No. 1107838 and No. 1408685. ARH was supported by NSF IGERT Fellowship Award No. 0801471.

\bibliography{myBibDatabase}
\end{document}